\documentclass[prl,superscriptaddress,twocolumn,showpacs,preprintnumbers,amsmath,amssymb]{revtex4}

\usepackage{graphicx}

\begin{document}

\title{Production and state-selective detection of ultracold, ground state RbCs molecules}

\author{Andrew J. Kerman}
\altaffiliation{Present address: MIT-Harvard Center for Ultracold Atoms, MIT, Cambridge, MA, USA}
\author{Jeremy M. Sage}
\author{Sunil Sainis}
\affiliation{Department of Physics, Yale University, New Haven, CT 06520, USA}
\author{Thomas Bergeman}
\affiliation{Department of Physics and Astronomy, SUNY, Stony Brook, NY 11794-3800, USA}
\author{David DeMille}
\affiliation{Department of Physics, Yale University, New Haven, CT 06520, USA}

\date{\today}

\begin{abstract}
Using resonance-enhanced two-photon ionization, we detect ultracold, ground-state RbCs molecules formed via photoassociation in a laser-cooled mixture of $^{85}$Rb and $^{133}$Cs atoms. We obtain extensive bound-bound excitation spectra of these molecules, which provide detailed information about their vibrational distribution, as well as spectroscopic data on the RbCs ground a$^3\Sigma^+$ and excited (2)$^3\Sigma^+$, (1)$^1\Pi$ states. Analysis of this data allows us to predict strong transitions from observed excited levels to the absolute vibronic ground state of RbCs, potentially allowing the production of stable, ultracold polar molecules at rates as large as $10^7$ s$^{-1}$.
\end{abstract}



\maketitle

Much experimental effort in recent years has been directed towards the production of polar molecules at ultracold temperatures \cite{buffer,stark}. Potential applications include quantum computation \cite{Qcomp}, study of highly correlated many-body systems \cite{condmat}, and sensitive tests of fundamental symmetries \cite{EDM}. Cooling of molecules, however, presents a significant challenge; their lack of closed optical transitions and expected inelastic collisional losses \cite{molinel} may preclude the use of robust laser and evaporative cooling techniques employed for many atomic species. Successful experiments so far have instead used either buffer-gas cooling \cite{buffer} or Stark slowing \cite{stark} to produce trapped polar molecules at temperatures down to $\sim$10 mK.

Another promising route involves the formation of heteronuclear diatomic molecules out of cold atoms \cite{wangstwa,ourRbCs,hetero}, which allows the success of atomic cooling techniques to be exploited to reach molecular temperatures down to the quantum-degenerate regime \cite{molBEC}. Until recently, however, these methods had been successfully applied only to homonuclear (and therefore non-polar) dimers \cite{homoion,homo,molBEC}. In our previous work, we demonstrated their extension to the heteronuclear case, and observed the formation of excited, polar RbCs$^\star$ molecules using photoassociation of ultracold, trapped $^{85}$Rb and $^{133}$Cs atoms \cite{ourRbCs}. We also predicted highly favorable rates of ground-state RbCs production; however, as in all schemes involving molecular formation in atomic collisions, the molecules are formed predominantly in highly excited vibrational levels. Such levels are likely to be unstable with respect to inelastic collisions \cite{molinel}, and to have small or vanishing electric dipole moments  (in the limit of small binding energy) \cite{jul03}. It is therefore desirable to transfer the molecules to their absolute rovibronic ground state.


In this Letter, we describe the direct detection of ultracold, ground-state RbCs molecules formed via photoassociation. This is accomplished using resonance-enhanced, two-photon ionization \cite{homoion,bagnato}, which also allows us to obtain extensive bound-bound molecular spectra. From these spectra we extract the level structure of both the ground a$^3\Sigma^+$ state (in which a sizeable fraction of our molecules are produced), and the excited (2)$^3\Sigma^+$ and (1)$^1\Pi$ states (to which we then resonantly excite them). We also infer from these observations exactly which a$^3\Sigma^+$ vibrational levels are most highly populated by photoassociation. Finally, based on our analysis we predict that the coupled (2)$^3\Sigma^+$ and (1)$^1\Pi$ levels that we excite during photoionization can be coupled to the absolute vibrational ground state of the molecule with a single additional photon. This should allow the production of stable, ultracold, polar molecules in this state at rates as large as $10^7$
s$^{-1}$.

\begin{figure}
\includegraphics[width=3.4in]{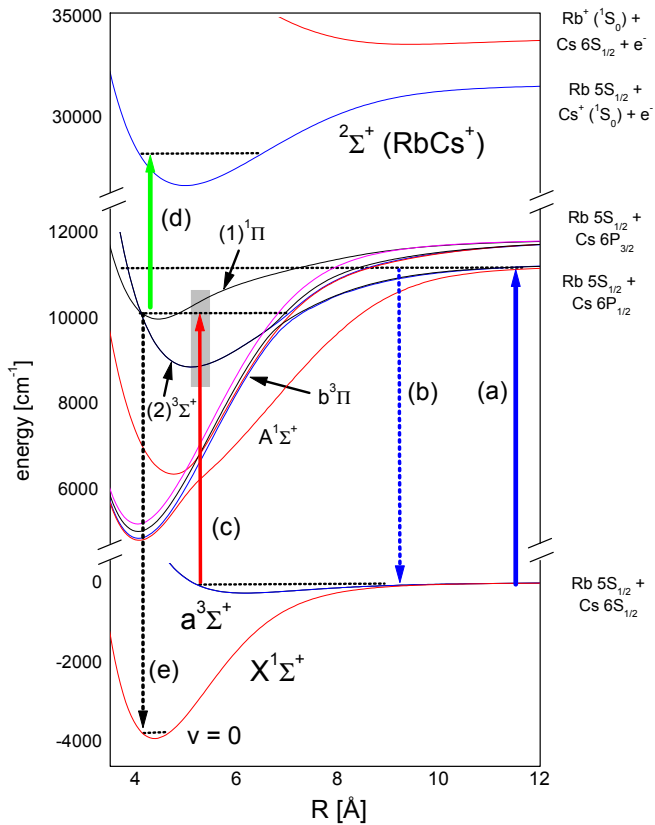}
\caption{(color online) Ultracold RbCs formation and detection processes. (a) Colliding ground state atom pairs are excited into weakly-bound levels of RbCs$^\star$. (b) excited molecules can decay into the a$^3\Sigma^+$ and X$^1\Sigma^+$ ground states; in this work we focus on the former. (c) These molecules are excited by a weak infrared laser pulse predominantly to the (2)$^3\Sigma^+$ and (1)$^1\Pi$ states (in the range shown by the shaded rectangle), and are then (d) ionized by an intense 532 nm pulse. (e) From coupled (2)$^3\Sigma^+$ and (1)$^1\Pi$ vibrational levels observed in this work, we predict that transitions to X$^1\Sigma^+(v=0)$ should be strong. }
\label{figure1}
\end{figure}

Fig. \ref{figure1} shows schematically the methods by which we produce and detect ground state RbCs molecules. Two colliding atoms absorb a photon and are promoted to a weakly-bound, electronically excited molecular level, a process known as photoassociation (PA) \cite{homoion}. They subsequently decay either to two free atoms or a ground-state molecule, and we detect these molecules using resonance-enhanced two-photon ionization (RE2PI) \cite{homoion}. A weak laser pulse drives a resonant, bound-bound transition from the a$^3\Sigma^+$ state to the (2)$^3\Sigma^+$ and/or (1)$^1\Pi$ states, after which the excited molecule is ionized by an intense, shorter-wavelength pulse. The energy of the ionizing photon is chosen to selectively form a RbCs$^+$ molecular ion (see Fig. \ref{figure1}), which we then detect using time-of-flight mass spectroscopy \cite{homoion}. Bound-bound molecular spectra are obtained by scanning the frequency of the first pulse, and monitoring the ion
signal.


The $^{85}$Rb and $^{133}$Cs atoms used for PA were collected and cooled in a dual-species, forced dark SPOT \cite{darkspot} magneto-optical trap (MOT) \cite{ourRbCs}. The atomic density $n$ and atom number $N$ were measured to be $n_{Rb}=7\times10^{11}$ cm$^{-3}$, $N_{Rb}=2\times10^8$, $n_{Cs}=1\times10^{12}$ cm$^{-3}$, $N_{Cs}=8\times10^8$ using two-color absorption imaging from two orthogonal directions as well as resonance fluorescence. The temperatures of both species were $\sim$75 $\mu$K, measured by time-of-flight absorption imaging. The PA transition was driven by a Ti:sapphire laser producing $\sim$500 mW around 900 nm, just below the lowest atomic asymptote Rb 5S$_{1/2}$ + Cs 6P$_{1/2}$ (Fig. \ref{figure1}). This beam was focused to a $e^{-2}$ waist size of $\sim$ 380 $\mu$m, producing an intensity more than sufficient to saturate the PA resonances used here. Its frequency was actively locked to the desired resonance using an optical spectrum analyzer, with the Rb trap laser as a reference.

The RE2PI laser pulses were both of $\sim$7 ns duration, and were separated in time by $\sim$10 ns. The first pulse had a tunable frequency in the near-infrared (IR), from $8350\rightarrow10650$ cm$^{-1}$, and a typical peak intensity of $3\times10^8$ W/m$^2$. It was generated using the output of a pulsed dye laser operating from $14000\rightarrow18400$ cm$^{-1}$ at 10 Hz with pulse energies up to $\sim$20 mJ, and a spectral linewidth $\le$ 0.05 cm$^{-1}$. This output was sent through a H$_2$ Raman cell (Light Age LAI 101 PAL-RC), which coherently produces additional frequency components offset by 4155.1 cm$^{-1}$ (the H$_2$ vibrational splitting). A dispersing prism was used to spatially separate the first or second Stokes (downshifted) Raman order, which was directed into the vacuum chamber. Its frequency was monitored using a wavemeter with 0.05 cm$^{-1}$ absolute accuracy. The second pulse, at 532 nm, had a typical intensity of of $6\times10^9$ W/m$^2$ and was derived from the doubled Nd:YAG laser used to pump the dye laser.


\begin{figure*}
\includegraphics[width=6.8in]{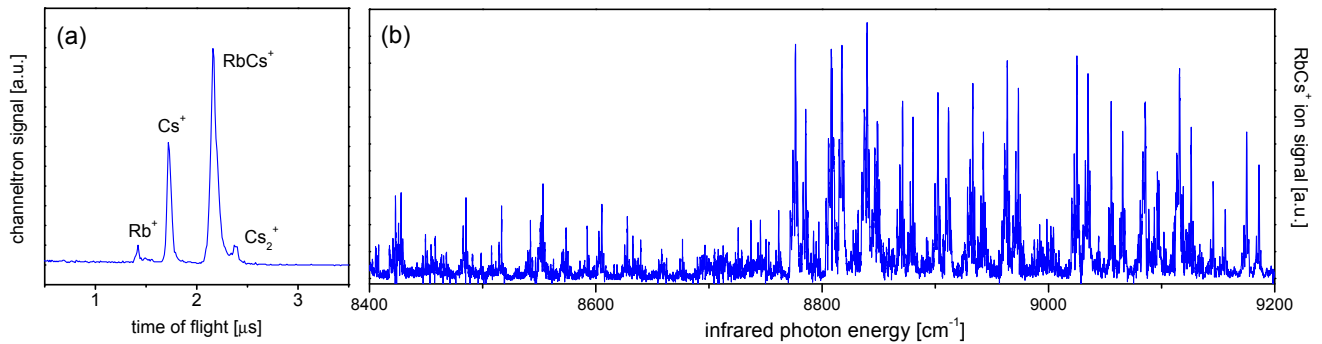}
\caption{Detection and spectroscopy of ultracold RbCs molecules. (a) Time-of-flight mass spectrum. The channeltron current, averaged over twenty shots, is plotted vs. delay time after the laser pulse. The PA laser is tuned to the RbCs$^\star$ level at $\Delta=-38.019$ cm$^{-1}$ having $\Omega=0^-,J=1$ \cite{ourRbCs}, which can decay only to a$^3\Sigma^+$ levels with $\Omega\in0^-,1$ and $J\in0,2$ \cite{singlet}. (b) Bound-bound RE2PI spectrum, showing a portion of the total scan range indicated by the shaded rectangle in Fig. \ref{figure1}. The strong progression beginning around 8770 cm$^{-1}$ is associated with the (2)$^3\Sigma^+$ excited state, whose $v=0$ level corresponds to the sudden onset of the series. The doubling of peaks occurs due to the spin-orbit splitting of its $\Omega=0^-,1$ components.} \label{figure2}
\end{figure*}

The ions produced by RE2PI were detected using a channeltron (Burle 5901) biased at -2 kV, $\sim$ 3 cm from the atoms. A second electrode on the opposite side of the atoms was biased at +2 kV. The channeltron current was digitized during a 2 $\mu$s interval after each laser pulse, resulting in a time-of flight mass spectrum like that shown in Fig. \ref{figure2}(a). The RbCs$^+$ mass peak was only observed if both Rb and Cs atoms were trapped, \textit{and} the PA laser was resonant with a suitable excited RbCs$^\star$ level. The Rb$^+$, Cs$^+$, and Cs$_2^+$ peaks in the spectrum arose primarily from the ionization of trapped Rb and Cs atoms, and Cs$_2$ molecules produced by the trapping light (that is, independent of the PA laser) \cite{nonres}.

To ensure that we detect RbCs molecules only in their a$^3\Sigma^+$ or X$^1\Sigma^+$ electronic ground states, the PA laser was extinguished $\sim$100 $\mu$s before each ionizing pulse, allowing any electronically excited molecules to decay. By increasing this delay time and monitoring the accompanying reduction in the RbCs$^+$ ion signal (due to ballistic flight of the RbCs molecules out of the ionization beam), we extracted an estimate of their kinetic temperature. From the observed decay time of the RbCs$^+$ signal ($\sim$10 ms) and the measured size of our ionization beams ($\sim$2 mm), we estimate this temperature to be $\sim$100 $\mu$K, comparable to the measured temperature of our atoms.

To understand which RbCs vibrational levels are populated by PA, we obtained bound-bound molecular spectra by scanning the frequency of the IR pulse, and a small segment of this scan is shown in Fig. \ref{figure2}(b). The intensities of the IR and 532 nm pulses were kept sufficiently low that neither by itself, nor the two in combination, produced RbCs$^+$ ions off-resonantly \cite{nonres}. The 532 nm pulse was chopped off for every other shot, and the mass spectra with and without it were subtracted for each frequency; this removed occasional contributions from multi-photon resonant processes driven purely by the IR pulse.

A vibrational progression is clearly evident in Fig. \ref{figure2}(b) beginning at $\sim$8770 cm$^{-1}$, consisting of a series of doublets, which we identify with transitions a$^3\Sigma^+\rightarrow(2)^3\Sigma^+$. The observed vibrational splitting of $\sim$33 cm$^{-1}$ is close to that predicted for the (2)$^3\Sigma^+$ state, and the onset of the series occurs close to the predicted minimum energy of that state \cite{RbCspot}. The doublet structure of each (2)$^3\Sigma^+$ level can be identified with the second-order spin-orbit splitting between its $\Omega=0^-,1$ components, observed here to be $\sim$9 cm$^{-1}$, nearly an order of magnitude larger than predicted \cite{RbCspot}.

\begin{figure}
\includegraphics[width=3.4in]{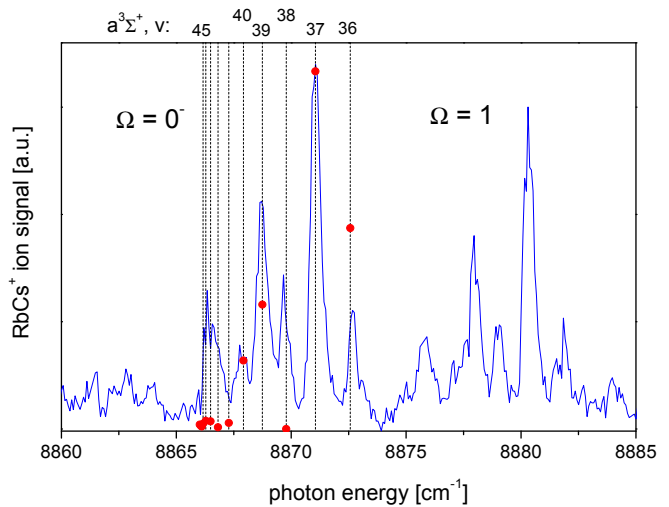}
\caption{Substructure of the a$^3\Sigma^+\rightarrow(2)^3\Sigma^+,v=4$ transition. The two sets of peaks arise from the $\Omega=0^-,1$ components of (2)$^3\Sigma^+,v=4$. The dashed lines show the predicted vibrational structure of the a$^3\Sigma^+$ state \cite{RbCspot}, and the solid circles indicate the predicted relative populations. The absolute position of this pattern was adjusted to match the observed features, which fixes the (2)$^3\Sigma^+$ level position. Note that the $\Omega=0^-,1$ components of a$^3\Sigma^+$ levels and the hyperfine/rotational substructure of both a$^3\Sigma^+$ and (2)$^3\Sigma^+$ are not resolved within the $\sim$0.05 cm$^{-1}$ linewidth of our laser.} \label{figure3}
\end{figure}

All of the strong doublets in Fig. \ref{figure2}(b) also exhibit a characteristic substructure, which arises from the vibrational levels of the ground a$^3\Sigma^+$ state. This is shown in detail in Fig. \ref{figure3}. The observed splittings agree well with those we predict for weakly-bound a$^3\Sigma^+$ vibrational levels from \textit{ab initio} potentials \cite{RbCspot}. In addition, since we observe most of these bound-bound transitions to be saturated, the relative heights of the features directly reflect the population distribution among the various a$^3\Sigma^+$ levels \cite{nodes}. This distribution agrees qualitatively with our calculations based on previous analysis of PA spectra \cite{ourRbCs} and the \textit{ab initio} a$^3\Sigma^+$ potential \cite{RbCspot}. Note that the inner turning points of these weakly bound a$^3\Sigma^+$ levels populated by PA nearly coincide with the minimum of the (2)$^3\Sigma^+$ state, as shown in Fig. \ref{figure1}; this should result in large Franck-Condon factors (FCFs) for excitation to low-lying levels of (2)$^3\Sigma^+$, and explains the ease with which we excite them. Finally, we note that the weaker peaks below 8770 cm$^{-1}$ in Fig. \ref{figure2}(b) exhibit the same a$^3\Sigma^+$ vibrational substructure, and arise from transitions from these levels to the coupled A$^1\Sigma^+$ and b$^3\Pi$ states.

Using the observed PA rate inferred from trap-loss measurements \cite{ourRbCs}, and the predicted FCFs for decay to a$^3\Sigma^+$ levels \cite{transdip} from our analysis of PA spectra \cite{ourRbCs}, we would expect RbCs
molecules to be formed in a$^3\Sigma^+(v=37)$ at a rate of $\sim10^7$ s$^{-1}$ in these experiments. However, we cannot yet quantitatively verify this prediction, due to a lack of knowledge of our absolute ionization efficiency for these molecules. Although the IR step of the RE2PI is observed to be saturated on most a$^3\Sigma^+\rightarrow(2)^3\Sigma^+,(1)^1\Pi$ resonances, the 532 nm step displays a more complex behavior; as its intensity is increased, broadened resonant features appear in both the Rb$^+$ and Cs$^+$ channels, indicating that multi-photon processes and possibly predissociation are involved. Based on the RbCs$^+$ signal corresponding to a saturated a$^3\Sigma^+(v=37)\rightarrow(2)^3\Sigma^+_0(v^\prime)$ transition, and the approximate gain and quantum efficiency of the channeltron, we infer that up to $\sim$250 ions are detected per pulse. Given the observed 10 ms time constant for decay of the ion signal, and the width of these bound-bound features ($\sim$ twenty times our laser linewidth \cite{linewidth}) this corresponds to a formation rate of a$^3\Sigma^+(v=37)$ molecules of $\sim5\times10^5$ s$^{-1}/E$, where $E$ is the ionization efficiency into the RbCs$^+$ channel.

Our bound-bound spectrum extends well beyond the predicted minimum of the (1)$^1\Pi$ state \cite{RbCspot}, and we indeed observe the onset of a more complex level structure near this energy. Clearly evident in this structure are strong perturbations due to off-diagonal spin-orbit interactions between (1)$^1\Pi$, (2)$^3\Sigma^+$, and b$^3\Pi$. These interactions are of crucial importance for our purposes, since an admixture of the (1)$^1\Pi$ state makes direct transitions to X$^1\Sigma^+$ possible, and allows us to identify a promising route for producing ultracold polar RbCs molecules in their stable absolute ground state X$^1\Sigma^+(v=0)$ at large rates. We have already observed that the a$^3\Sigma^+$ levels most highly populated by PA can be transferred to any of the coupled (2)$^3\Sigma^+$, (1)$^1\Pi$ levels up to 10650 cm$^{-1}$. In Fig. \ref{figure1} we see that low-lying (1)$^1\Pi$ levels should have large FCFs for transitions directly to X$^1\Sigma^+(v=0)$ (as indicated by the arrow marked ``(e)" in Fig. \ref{figure1}), since the vibrational turning points nearly coincide. In fact, from preliminary analysis of our spectra we predict that these FCFs are $\sim$0.1 for a range of low-lying (1)$^1\Pi$ vibrational levels; thus, a stimulated transition would require only a modest intensity (in this case at about 715 nm) and could make efficient transfer from a$^3\Sigma^+(v=37)\rightarrow$ X$^1\Sigma^+(v=0)$ possible \cite{gusym}. It should be pointed out, however, that at present our a$^3\Sigma^+(v=37)$ molecules are distributed among several hyperfine/rotational levels, and thus only a fraction of them could be transferred in this way. The use of a spin-polarized atomic sample in an optical trap, as well as hyperfine-resolved PA to an $\Omega=1$ level \cite{ourRbCs} could potentially solve this problem, and narrow-band pulsed lasers could be used to selectively populate a single hyperfine/rotational component of X$^1\Sigma^+(v=0)$.


In summary, we have produced ultracold, ground-state RbCs molecules using photoassociation. We have detected these molecules using resonance-enhanced two-photon ionization, and obtained bound-bound molecular spectra of the ground a$^3\Sigma^+$ and excited (2)$^3\Sigma^+$ and (1)$^1\Pi$ states. From these spectra we have obtained information about the distribution of ground-state vibrational levels in which ultracold molecules are produced via photoassociation; further, we have identified a promising route in which the a$^3\Sigma^+$ vibrational levels most highly populated by photoassociation can be transferred directly to the absolute ground state of the molecule, and we have demonstrated the first step of this process. With the addition of a laser to resonantly transfer the excited (1)$^1\Pi$ molecules to X$^1\Sigma^+(v=0)$, we should be able to produce large samples of collisionally stable, ultracold polar molecules.

We acknowledge support from NSF grant EIA-0081332, the David
and Lucile Packard Foundation, and the W.M. Keck Foundation. T.B. acknowledges funding from the
U.S. Office of Naval Research.


\begin{thebibliography}{}
\bibitem{singlet} We have also observed features associated with X$^1\Sigma^+$ molecules (using $\Omega=0^+,1$ PA resonances); however, these features were weak and hard to assign, due to small ionization FCFs.
\bibitem{transdip} Here, the atomic transition dipole is assumed; this should be a good approximation, since the decay occurs almost purely at long range.
\bibitem{nodes} There are several sharp minima in the strength of the observed lines, e.g. in Fig. \ref{figure2}(b) near 9000 cm$^{-1}$ and 9150 cm$^{-1}$. These lines are not saturated, and are suppressed due to an overlap between a node in the excited vibrational wavefunction and the inner turning points of the initial a$^3\Sigma^+$ levels.
\bibitem{nonres} A fraction of the Cs$^+$ ions, however, were produced from RbCs molecules via off-resonant multi-photon ionization \cite{bagnato} by the 532 nm pulse.
\bibitem{linewidth} In correcting for the observed linewidth we have assumed that it is dominated by ground-state hyperfine/rotational structure, and that the excited state structure is smaller than the laser linewidth and can be neglected; this should be a good assumption given the small $\Omega=0$ hyperfine splitting, and the limited number of accessible rotational levels.
\bibitem{gusym} Note that this type of route to the X$^1\Sigma^+$ state starting from a$^3\Sigma^+$ levels is forbidden in the homonuclear case, due to the additional g/u electronic symmetry.
\end{thebibliography}
\end{document}